\documentclass[natbib]{svjour3}                     
\smartqed  
\usepackage{graphicx}
\usepackage{hyperref}
\usepackage{times}
 \usepackage{aps-bibstyle}  
\newcommand{\hii}{H{\small\rm II}\relax} 
\newcommand{\beq}{\begin{equation}}
\newcommand{\eeq}{\end{equation}}
\def\blfootnote{\xdef\@thefnmark{}\@footnotetext}

\newcommand\alp{\mbox{$\alpha$}}

\newcommand\arcmin{\hbox{$^\prime$}}

\newcommand\solar{\mbox{$_{\normalsize\odot}$}}

\newcommand{\AmS}{{\protect\the\textfont2
  A\kern-.1667em\lower.5ex\hbox{M}\kern-.125emS}}
\newcommand{\lsim}{\ \raise
-2.truept\hbox{\rlap{\hbox{$\sim$}}\raise5.truept\hbox{$<$}\ }}
\newcommand{\gsim}{\ \raise
-2.truept\hbox{\rlap{\hbox{$\sim$}}\raise5.truept\hbox{$>$}\ }}
\newcommand{\simsim}{\ \raise
-2.truept\hbox{\rlap{\hbox{$\sim$}}\raise5.truept\hbox{$\sim$}\ }}

\newcommand{\aap}{{Astron. Astrophys.}}
\newcommand{\apj}{{Astrophys. J.}}

\newcommand{\apjl}{{Astrophys. J. Lett.}}
\newcommand{\mnras}{{MNRAS}}
\newcommand{\araa}{{Ann. Rev. Astron. Astrophys.}}
\newcommand{\aapr}{{Astron. Astrophys. Rev.}}
\newcommand{\aj}{{Astron. J.}}
\newcommand{\pasp}{{Publ. Astron. Soc. Pacific}}
\newcommand{\sci}{{Science}}
\newcommand{\apjs}{{Astrophys. J. Suppl. Ser.}}
\newcommand{\apss}{{Astrophys. Space Sci.}}
\begin{document}

\title{
Low-mass pre--main-sequence stars in the Magellanic Clouds
}


\author{
Dimitrios~A.~Gouliermis
}


\institute{
              Dimitrios A. Gouliermis \at
              Max Planck Institut f\"{u}r Astronomie,  K\"{o}nigstuhl 17, D-69117 Heidelberg, Germany \\
              \email{dgoulier@mpia-hd.mpg.de
              }           
}

\date{Received: September 27, 2011 / Accepted: February 7, 2012}

\titlerunning{ 
Low-mass pre--main-sequence stars in the Magellanic Clouds
} 
\authorrunning{ 
D. A. Gouliermis
} 

\maketitle

\begin{abstract}
The stellar Initial Mass Function (IMF) suggests that stars with sub-solar mass form in very large numbers.
Most attractive places for catching low-mass star formation in the act are young stellar
clusters and associations, still (half-)embedded in star-forming regions. The low-mass stars 
in such regions are still in their pre--main-sequence (PMS) evolutionary phase, 
i.e., they have not started their lives on the main-sequence yet. The peculiar nature
of these objects and the contamination
of their samples by the fore- and background evolved populations of the Galactic disk  
impose demanding observational techniques, such as X-ray surveying and optical spectroscopy
of large samples for the detection of complete
numbers of PMS stars in the  Milky Way. The Magellanic Clouds, the metal-poor companion
galaxies to our own, demonstrate an exceptional star formation activity. The low extinction and 
stellar field contamination in star-forming regions of these galaxies imply a more
efficient detection of low-mass PMS stars than in the Milky Way, but their distance 
from us make the application of the above techniques unfeasible.  
Nonetheless, imaging with the {\sl Hubble Space Telescope} within 
the last five years yield the discovery of solar and sub-solar 
PMS stars in the Magellanic Clouds from photometry alone. Unprecedented numbers  of such objects are identified 
as the low-mass stellar content of star-forming regions in these galaxies, changing completely 
our picture of  young stellar systems outside the Milky Way, and extending 
the extragalactic stellar IMF below the persisting threshold of a few solar 
masses. This review presents the recent developments in the investigation of the PMS stellar 
content of the Magellanic Clouds, with special focus on the limitations by single-epoch 
photometry that can only be circumvented by the detailed study of the observable behavior 
of these stars in the color-magnitude diagram. The achieved characterization of the low-mass PMS stars 
in the Magellanic Clouds allowed thus a more comprehensive understanding 
of the star formation process in our neighboring galaxies. 

\end{abstract}

\section{Introduction}\label{s:intro}

 {The Large and Small Magellanic Clouds (LMC, SMC) are the closest
undisrupted neighboring galaxies to our own. They are characterized by 
their low metal abundance, which is 2.5 and 5 times respectively lower than Solar 
\citep[][]{westerlund97, luck98, venn99}, and their dust-to-gas ratios, which vary 
from 2 up to 10 times lower than in the Milky Way \citep{stanimirovic00, gordon03}. 
Their metallicities suggest that the environments
of the Magellanic Clouds (MCs) are closer to that of the early universe
during the epoch of its peak star formation at $z\sim$~1.5 \citep{madau96, 
pei99}. They are, therefore, the best local templates of primitive star formation.}
As such, both the MCs offer an outstanding variety 
of young stellar clusters and associations, the age and Initial Mass 
Function (IMF) of which become very important sources of information on 
their recent star formation. Young star clusters and associations contain the richest 
samples of bright OB-type stars in a galaxy. Consequently our knowledge on 
the young massive stars of the MCs has been collected from 
photometric and spectroscopic studies of such systems \citep[see, e.g.,][and references 
therein]{massey06}.  {In particular, recent spectroscopic studies in the 
optical \citep[see, e.g.,][]{evans06, hunter09, evans11} and infrared 
\citep[e.g.,][]{bonanos10, sewilo10} provide a unique insight of the properties 
of young massive stars in the MCs.} However, a more comprehensive picture of 
young stellar systems in these galaxies has emerged 
when {\sl Hubble Space Telescope} ({\sl HST}) observations revealed that 
star-forming regions of the MCs host large numbers of faint 
pre--main-sequence (PMS) stars.

 {The youngest star clusters and associations with ages $<~10$~Myr are 
embedded in bright \hii\  regions, comprising high-mass stars on the main-sequence 
(MS) as well as intermediate- and low-mass young stellar objects (YSOs),
i.e., stars in the earliest stages of their evolution. There are two principal kinds of 
YSOs, protostars, denoting stars in their early stages of formation, and PMS 
stars \citep[see, e.g.,][]{lada00}. According to the classification by \cite{robitaille06}, 
analogous to the traditional scheme of \cite{lada87}, stage 0-I  YSOs  have ages 
$<$ 0.1 Myr and are characterized by infalling envelopes. Stage II YSOs have ages 
between 0.1 and a few Myr and demonstrate thick disks.  The more evolved stage III 
YSOs  have optically thin disks or no disks at all. PMS stars are usually associated
with YSOs of stages II-III \citep{lada99}.} The high-mass (OB-type) 
MS stars and intermediate-mass PMS populations  (with typical example the 
Herbig Ae/Be stars; \citealt{ww98}) are the direct signature of 
the youthfulness of their hosting stellar systems, but they represent only
the latest phases of star formation. On the other hand, the low-mass 
PMS stars of these systems preserve a record of the complete recent star formation 
history of the star-forming region over long periods, because their evolution is extremely 
slow and can last up to many tens of Myr\footnote{Typical contraction time for a 1~M{\solar} 
PMS star is 50~Myr and for a 0.5~M{\solar} PMS star 200~Myr \citep[][]{karttunen07}.}. Indeed, 
the star formation history of star-forming regions in the Galaxy is usually 
constrained by the study of their PMS stars \citep[see e.g.,][]{preibisch02, sherry04, briceno07, 
reipurth08a, reipurth08b, preibisch11}. However, the samples of these stars are significantly 
contaminated by the field of the Galactic plane, and therefore various observational techniques 
are developed for the correct identification of faint PMS stars, and in particular T~Tauri stars 
(TTS; \citealt{am89}) as the typical paradigm of such stars, in star-forming regions of the Milky Way. 

PMS stars exhibit periodic fluctuations in light that indicate large, rotating star\-spots. 
Therefore, optical variability surveys are used for the collection of large samples
of such stars in the Galaxy. Prominent optical emission lines, as well as  X-ray emission, 
are thought to stem from chromospheric heating. Spectroscopy is thus quite 
useful for the detection of Galactic PMS stars, and X-ray observations 
have been especially useful in locating PMS objects in dense molecular cloud 
environments.
Moreover, since these stars are still accreting material through circumstellar 
disks, deep UV and H{$\alpha$} imaging is also applied for accretion Ê
studies through the detection of excess emission in these wavelengths.
Excess in near-IR wavelengths is a positive signature of disks around young 
stars, and therefore PMS stars are also identified through their near-IR colors.
 {Other techniques for the investigation of Galactic PMS stars involve Lithium 
abundance measurements and dating of the stars from Li depletion 
\citep[e.g.,][]{jeffries05, mentuch08},  far- and mid-IR imaging \citep[e.g.,][]{merin08,
baume11}, and interferometry for the study of circumstellar disks properties 
\citep[e.g.,][]{ragland11}.} While such techniques are proven to be profitable in studying 
faint Galactic PMS stars, at the distance of the MCs 
($\sim$~50~-~60~kpc) finding such stars can only be achieved through imaging at the 
angular resolution and wide-field coverage facilitated with {\sl HST}. In addition, 
observational and technical limitations do not allow the successful application
of techniques such as multi-object spectroscopy, time-dependent imaging for
variability studies, or X-ray surveys, among others, and thus investigators have 
to rely on single-epoch photometry alone for detecting and characterizing faint PMS 
stars in the MCs. Nevertheless, considering the aforementioned characteristics of
PMS stars, the positions of these stars in the color-magnitude diagram (CMD)
can be misleading and introduce biases that affect their interpretation.

This problem can be overcome by the successful distinction of the PMS 
stars from their surrounding field populations, and the thorough study of 
the behavior of these stars in the CMD through modeling of their 
physical properties. To this end, a successful study of significant numbers of 
such stars in star-forming regions of the MCs concern the complete 
understanding of observational and physical factors that affect the
measured magnitudes and colors of these stars, in order to achieve 
a comprehensive assessment of their photometric behavior 
as members of a young stellar cluster, and the accurate determination of the 
IMF and the age of this cluster. This paper is an overview of all studies focused on the investigation of
PMS stars in the MCs from {\sl HST} photometry.  {Based on the lack of previous 
concrete data on PMS stars in the MCs, the fundamental assumption, i.e., the working 
hypothesis, of such studies is that the observable behavior of low-mass PMS stars in 
the MCs does not differ from that of  their Milky Way TTS 
counterparts \citep[see, e.g.,][]{gouliermis10}. }

This review is constructed as 
the following. In \S\ref{s:pmsdetect} I  present the recent detections of faint PMS stars in various star-forming 
regions of both the MCs with {\sl HST}. The optical CMD of a typical young cluster in 
the MCs is discussed in \S\ref{s:cmdposdisl}, where I also give a detailed account of the 
biasing factors that affect the observed CMD-positions of the detected
PMS stars. Calculations of PMS evolutionary models, and their transformation 
to the observable plane, i.e., the CMD, are extremely useful 
for the accurate translation of magnitudes and colors into masses 
and ages of PMS stars. Methods for this translation are discussed in \S\ref{s:mod2dat}. 
In the following sections an account of statistical analyses of the rich samples of 
PMS stars found with {\sl HST}  in MCs star-forming regions is given.
In particular the determination of the stellar IMF in MCs young clusters from their
PMS populations is described in \S\ref{s:imflh95}, and methods for the assessment 
of the age of PMS clusters, and their intrinsic age-spreads are presented in \S\ref{s:timesf}.
Finally, conclusive remarks are given in \S\ref{s:conclrem}.

\section{Detection of PMS stars in the MCs}\label{s:pmsdetect}

\subsection{Early detections}

In the Galaxy earlier investigations  confirmed the
existence of faint PMS stars in a variety of star-forming environments,
from open clusters and associations, such as, e.g., NGC~6611 \citep{hillenbrand93}, 
Upper Scorpius \citep{preibisch99}, and Orion OB1, \citep{sherry04, briceno05}
to compact massive starbursts (e.g., NGC~3603, \citealt{brandl99}; \citealt{stolte04}). 
Pioneering studies on PMS stars in the MCs
were focused mostly on the starburst of 30~Doradus in the LMC. 
Historically, \cite{brandl96} were the first to detect 108 ``extremely red sources'', 
which are most likely PMS stars of low or intermediate mass from adaptive optics near-IR 
imaging of the central region of 30~Dor. They constructed the IMF of these stars 
down to $\sim$~3~M{\solar}.  \cite{sirianni00}, using imaging from the Wide-Field
Planetary Camera~2 ({\sl WFPC2}) onboard {\sl HST}, detected in the same 
region stars down to 1.35~M\solar\ and they found that the red population in the 
CMD is well traced by PMS isochrones. \cite{brandner01} 
observed again the area of 30~Dor with the Near Infrared Camera and Multi-Object Spectrometer 
({\sl NICMOS}) onboard {\sl HST} and they found no evidence for a lower 
mass cutoff in the IMF for PMS stars, in contrast to what was earlier reported. 
These studies revealed for the first time the PMS stellar content of the Tarantula nebula 
starburst. However, the region of 30~Dor in known to suffer from significant stellar crowding 
and high extinction, factors which limit the detection of complete samples of low-mass PMS
stars to about 2~M{\solar} \citep{romaniello06}. 

In contrast, less crowded clusters and associations, embedded in typical \hii\  
regions of the MCs, do not suffer severely from these factors, and thus faint PMS 
stars are easier detectable in such systems, proving them to be important hives of 
low-mass PMS populations.  \cite{brandner99} reported the detection  
of about 150 objects with excess emission in the SMC association NGC~346 with 
observations from {\sl HST~NICMOS}.  These authors explained this observation as the 
detection of candidate PMS stars with masses between 1 and 2 M{\solar}. 
Other early PMS studies in the MCs were conducted on the surrounding area 
of the supernova remnant SN~1987A in the LMC. A small number of 492 stellar 
sources was identified as 1~-~2~M{\solar} field PMS stars based on their 
conspicuous H$\alpha$ excess \citep{panagia00}. Subsequent studies in star-forming regions 
less crowded than 30~Dor are significantly benefitted by the angular resolution and wide-field 
coverage provided by {\sl HST}. 

The discovery of sub-solar PMS stars  in a star-forming region of the MCs less crowded 
than 30 Dor  was originally made with the use of archival imaging  taken with {\sl WFPC2} 
of the LMC stellar association LH~52 \citep{gouliermis06a}. 
The locations of the detected PMS candidates in the optical CMD
were found to be in excellent agreement with those of TTS of $M$~\lsim~2~M{\solar} in 
the Galactic association Ori OB1 \citep{sherry04, briceno05}.  
 {The WFPC2 images of LH~52 provided the first discovery of 
such stars in LMC star-forming regions, with the identification of about 500 sources 
down to $\sim$~0.3~M{\solar}.} However, this sample being limited by 
incompleteness did not allow a thorough statistical investigation.
Nevertheless, this study demonstarted that low-mass PMS stars can be directly 
identified in the CMD from photometry in $V$- and $I$-equivalent 
filters, providing that the local background field of the galaxy has been adequately  
observed. Subsequent deep imaging with the Wide-Field Channel (WFC) of the  
Advanced Camera for Surveys ({\sl ACS}) onboard {\sl HST} of the young clusters NGC~602 in the 
SMC and LH~95 in the LMC, and the bright star-forming regions N66 (NGC~346) in the SMC and 
N11 in the LMC, provided an outstanding insight of the low-mass stellar population in such regions.
I present these findings in the following section.

\begin{figure*}[t!]
\centerline{\includegraphics[clip=true,width=0.775\textwidth]{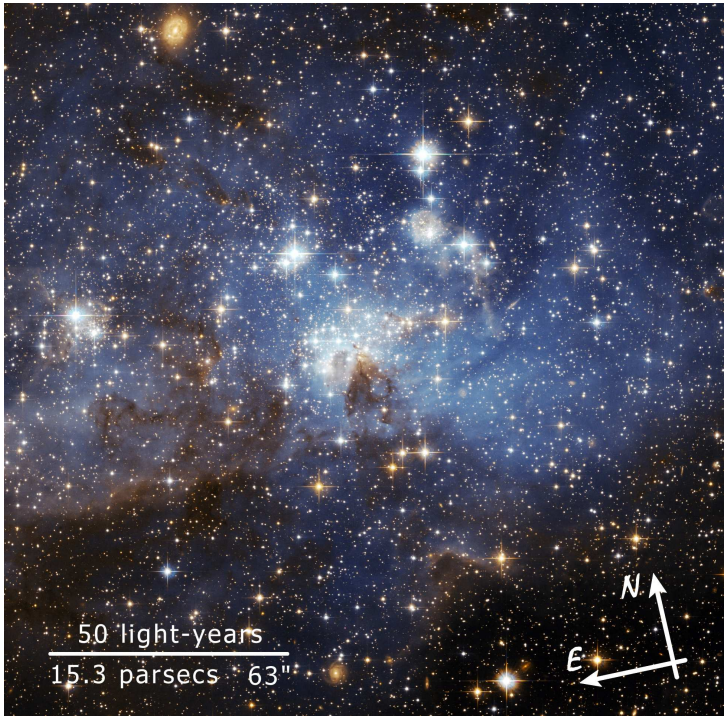}}
\caption{Color-composite image from ACS/WFC observations in the filters
F555W and F814W ($V$- and $I$-equivalent) of the LMC star-forming
region LH~95/N64. These observations revealed an outstanding number 
of low-mass infant PMS stars coexisting with young massive giant stars. The 
smooth continuum in the image is due to the diffuse H$\alpha$ emission.
Image credit: NASA, ESA and D. A. Gouliermis (MPIA). Acknowledgments: 
Lars Lindberg Christensen (ESO/ESA/ST-ECF), Davide de Martin 
(ESA/Hubble).\label{f:prpic}}
\end{figure*}

\subsection{PMS stars in young star clusters of the MCs}\label{s:pmsinmcs}

The photometric study with {\sl ACS}/WFC imaging of the association NGC~346 and 
its surrounding field, embedded in the \hii\  region LH{\alp}~115-N66 
\citep[in short N66;][]{henize56}, the brightest 
in the SMC, led to the discovery of an extraordinary number of PMS stars in 
its vicinity \citep{nota06, gouliermis06b}. Almost 100,000 stars in various 
evolutionary stages were detected over a magnitude range 
11~\lsim~$m_{\rm 555}$ \lsim~28, in three overlapping WFC 
pointings, covering a field about 5\arcmin\ ($\sim$~85~pc) wide. 
The bright MS stars of the association have been the subject 
of thorough investigation of massive stellar evolution in low metallicities
\citep{massey89, walborn00, evans06}. The {\sl ACS} observations demonstrated that 
the `evolved' bright MS stellar content is extended to the PMS regime at  fainter 
luminosities. The majority of the low-mass PMS stars in the region NGC~346/N66 
was found to be mostly concentrated in the association and several individual 
small subclusters in the vicinity of the \hii\  region \citep{sabbi07, 
hennekemper08, schmeja09}. 

The young SMC cluster NGC~602 is associated with the ring-shaped \hii\  
region N90  \citep{henize56}, located in the wing of the SMC. The photometry of a single 
{\sl ACS}/WFC
pointing of the system revealed that this cluster is characterized 
by no more than $\sim$~750 faint PMS stars (with $m_{\rm 555}$~\gsim~22.5), which are 
found to be well concentrated toward the center of the \hii\  ring \citep{carlson07,
schmalzl08, cignoni09}. 
The stellar population of the young cluster is known to comprise massive 
hot stars \citep[e.g.][]{battinelli98, massey00}, with the 
ten brightest  having spectral types between O6 and B5
\citep{hutchings91}. They belong to a sample of over 60  
MS stars, with $m_{\rm 555}$~\lsim~22.5, which are 
identified as members of the cluster with {\sl HST}. 
A small fraction of the 
faint PMS population of the region NGC~602/N90 is found in very compact 
concentrations, embedded in the dusty rim at the edge of the \hii\  cavity. Some of these
concentrations coinside with bright sources, and almost all of them appear very bright in
the mid-IR, as revealed from Spitzer/IRAC imaging of the region \citep{gouliermis07a, 
carlson11}, demonstrating their youthfullness.

The large \hii\  region LH{\alp}~120-N11 \citep{henize56}, in short N11, in the LMC 
is characterized by a central hole, surrounded by seven distinct bright nebulae 
and filaments. It comprises four known stellar associations, named LH~9, LH~10, LH~13 
and LH~14 \citep{luckehodge70}. The central cavity of the region is presumably evacuated 
by the OB stars in LH~9, located almost at the center of N11. 
Archival {\sl ACS} photometry complemented by archival IR data from {\sl Spitzer 
Space Telescope} revealed a large population of PMS candidates with masses in the range
1.3~-~2~M{\solar} and ages between  2 and 10~Myr \citep{vallenari10}. Embedded YSOs 
with ages of about 0.1~-~1~Myr are found to be intermixed with the 
PMS stars. The spatial distribution of all these young sources, compared with previous 
observations at different wavelengths and with the distribution of OB and candidate 
Herbig Ae/Be stars, showed that they are the product of clustered star formation, which is 
a  long-lasting process in N11.

\begin{figure*}[t!]
\centerline{\includegraphics[clip=true,width=\textwidth]{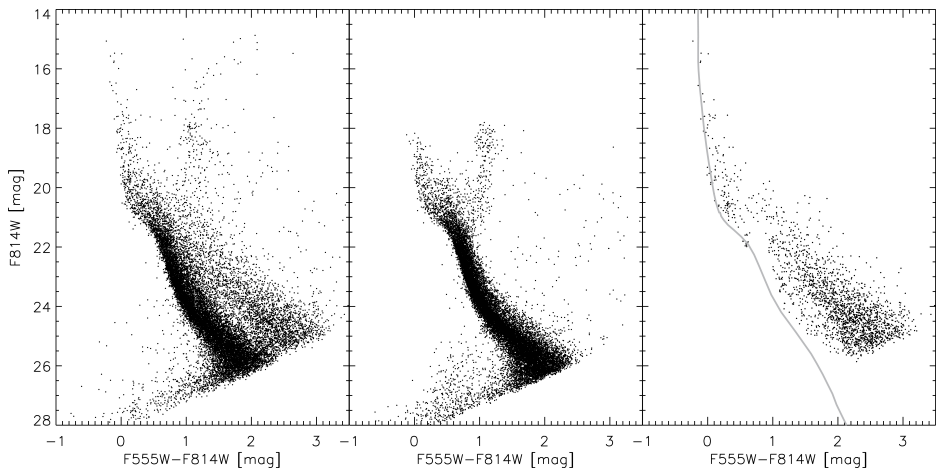}}
\caption{The F555W$-$F814W, F814W ($VI$-equivalent) CMDs of the stars detected with
{\sl ACS}/WFC within {\sl HST} Program GO~10566 in the young LMC cluster LH~95 
and its nearby background field \citep{gouliermis07b}. {\sl
Left:} The CMD of all stars found in the observed pointing of the system.
{\sl Middle:} The corresponding CMD of the control field, located about 5\arcmin\ 
($\sim$~75~pc) west of the system. The comparison of these CMDs exhibits the
differences in stellar content of the two observed regions, and
demonstrates the richness of PMS stars in the cluster region, as it is 
emphasized by the complete lack of such stars in the LMC field. 
{\sl Right:} The CMD of the main area of the system, only for stars which 
are selected as members of the cluster, after the contamination of the LMC 
field has been statistically subtracted. The ZAMS from the Padova grid of evolutionary
models \citep{girardi02} is overlaid for guidance. PMS evolutionary models 
suggest that the observed PMS population covers stellar masses down to 
$\sim$~0.2~M{\solar}. 
\label{f:lh95-cmd}}
\end{figure*}

The successful detection of the PMS population of a young star 
cluster down to the smallest possibly detectable mass  at the distance of the MCs 
requires high sensitivity and resolving efficiency in a wide-field coverage. Such 
observations naturally call for {\sl HST}, and in the optical regime 
{\sl ACS}/WFC with its unique combination of sensitivity, resolution and coverage, is the most 
appropriate camera even in the post-{\sl WFC3} era. In order to 
eliminate the confusion in the detectability of the faintest stars, loose clusters should be targeted, 
and complementary 
observations of a carefully  selected control-field should also be performed. 
Such a program designed to obtain {\sl ACS} images of the young LMC cluster LH~95 
and its nearby general field led to the deepest LMC observations ever taken with {\sl HST} 
\citep[][see also Fig.~\ref{f:prpic}]{gouliermis07b}. These 
observations allowed the construction of the CMD 
of the cluster in unprecedented detail and its accurate 
decontamination from the average LMC stellar population (Fig.~\ref{f:lh95-cmd}). 
While N64, the \hii\  region where LH~95 is embedded, represents a rather 
modest star-forming region, the photometry of these images, with a detection 
limit of $m_{\rm 555}$~\lsim~28 (at 50\% completeness), revealed more than 
2,500 PMS stars with masses down to $\sim$~0.2~M{\solar} in a single WFC 
pointing on the system.

\section{The optical CMD of PMS stars: Biasses in its interpretation}\label{s:cmdposdisl}

The CMD of LH~95 shown in Fig.~\ref{f:lh95-cmd} (right panel) represents a typical
optical CMD of a young star cluster in the MCs. It demonstrates that the early-type massive 
stellar members of such clusters are well aligned with the main-sequence down to about 
$m_{\rm 555} \simeq 21$, which corresponds roughly (depending on the model) to PMS stars of 
about 2~M{\solar}. In fainter magnitudes {\sl all} the stars of the cluster diverge from 
the main-sequence. They occupy the PMS part of the CMD, as is the case for typical star-forming 
regions in the Milky Way \citep[see, e.g.,][and contributions therein]{reipurth08a, reipurth08b}. 
These faint PMS stellar members of young star clusters spend extremely long periods of 
time remaining in this phase of their evolution.  {Indicatively, depending on the evolutionary
models, a 2~M{\solar} PMS star would need $\sim$~30~Myr for hydrogen ignition to start, while
this timescale for a  0.2~M{\solar} star becomes $\sim$~100~Myr.} In addition, their exceptionally large numbers 
cover  an important portion of the total mass of the cluster. As a consequence faint PMS stars 
posses crucial information about the age and the low-mass IMF of their hosting clusters. However, 
in order to extract this information, it is necessary for investigators to first disentangle their individual 
properties, i.e., their individual mass and age from their CMD positions alone. 

This task can only be achieved with the correct use of accurate theoretical PMS evolutionary 
models, because the use of different models can affect significantly the interpretation of the observed
CMDs.  Nevertheless, the simple direct ``translation'' of magnitudes and 
colors into masses and ages, as is usually the case for evolved stars, only accounting for photometric
errors and reddening, cannot be applied in the case of such faint PMS populations, as it will not be
reliable. The reason lays on the peculiar emission behavior of PMS stars and observational constrains 
that affect significantly the true positions of these stars in the CMD, introducing important biasses 
that must be first taken into account in the interpretation.  {In particular, a number of factors 
that can bias the age determination and  the assessment of any age spread among PMS stars 
have been widely discussed in the literature. Two nominal studies on the matter are those by 
\cite{hartmann2001}, who discuss  a number of biases that could produce a spread of low-mass 
PMS stars in the CMD without requiring a genuine spread in ages, and by \cite{hillenbrand2009}, 
who presents an overview of the age dating methods available for young sub-solar mass stars.
Factors that bias the CMD positions of PMS stars are discussed here in \S~\ref{s:cmdpmsbrd}.}

\begin{figure*}[t!]
\centerline{\includegraphics[clip=true,width=\textwidth]{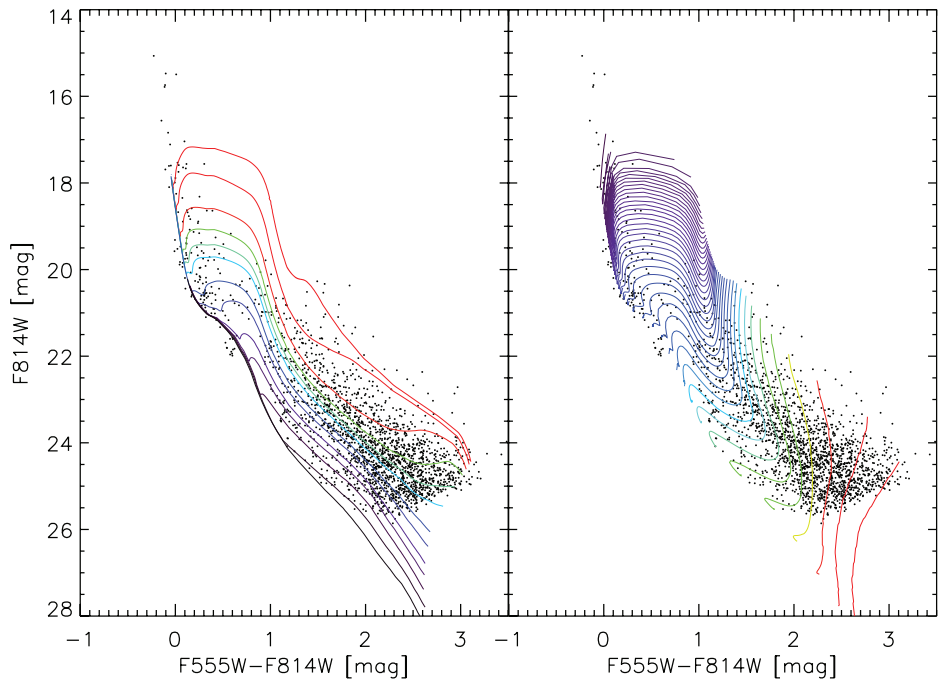}}
\caption{The F555W$-$F814W, F814W CMD of the stellar population of LH~95 
with evolutionary models calculated with the {\sc franec} code \citep{chieffi89, 
deglinnocenti08} for the LMC metallicity ($Z=0.01$) overlaid. {\sl Left:} PMS 
isochrones covering ages 
from 0.5 (red line) to 100~Myr (black line) from the {\sc franec} family of models. {\sl Right:} 
Evolutionary tracks for PMS stars covering masses from 0.2 (red line) to 
6~M{\solar} (black line) from the same family of models. The theoretical models
are converted into observable colors and magnitudes for the {\sl ACS} filter 
system in terms of use of model atmospheres for cool stars \citep[][see also~\S\ref{s:obsmodl}]{dario10}. 
Models are appropriately corrected for distance and average extinction. \label{f:cmdmods}}
\end{figure*}

\subsection{The effect of the theoretical PMS evolutionary models}\label{s:pmsmodel}

There are several different published calculations of PMS evolutionary 
models (see, e.g., \citealt{hillenbrand04} for an overview), which are 
constructed based on different assumptions, physics, and methods. Indicative 
references for seven families of widely used PMS evolutionary models are 
these published by \cite{swenson94}, \cite{dantonamazzitelli97}, \cite{baraffe98}, 
\cite{siess00}, \cite{pallastahler99}, \cite{yi03}, and \cite{deglinnocenti08}, updated by
\cite{tognelli11}. \cite{hillenbrand08} assess the 
systematic trends between the various sets of tracks by considering the 
predictions for some fiducial stars of given temperature and luminosity.  
These authors found that for sub-solar stars, systematic effects between 
the tracks are observed at the level of 0.75 dex at the youngest ages.
As a consequence, cluster age estimates are strongly dependent on 
the adopted set of PMS evolutionary theory.  There is a better agreement, 
particularly towards older ages, for solar-type stars.  {In a recent study 
of the ability of current evolutionary models in recovering the observed properties 
of PMS stars, \cite{gennaro11} used 25 PMS objects with well-measured 
dynamical masses, sixteen belonging to binary systems. These authors  found that while 
the masses of PMS stars in eclipsing binaries are generally well recovered within 
10\%, for one third of the binary systems, the ages derived for the two components 
are not in agreement. This implies an inefficiency of current PMS models in 
recovering precise stellar ages or alternatively that undetected systematics partly affect 
the interpretation of PMS binary observations.}

 {For all families of PMS models with ages younger than 50-80~Myr, high-mass 
stars are predicted 
to be older than low-mass stars in the same clusters \citep{hillenbrand2009}. 
While this effect has often been ascribed to the influence of the {\sl stellar birthline} \citep{stahler83, 
hartmann97}, it can also be due to the effect of magnetic fields, stellar rotation, or 
disk accretion. \cite{hillenbrand08} note that the trend of age with mass seems 
to persists longer than the influence of birthline effects is expected to last. In 
line to this, the recent models by \cite{baraffe09} suggest that for low-mass stars 
the concept of stellar birthline has no significance, because the first appearance of 
these stars in the HR diagram is random and depends on the accretion history. 
This can mimic an age dispersion.} \cite{darioONCII} in their recent study of the 
Orion Nebula Cluster (ONC) assign this mass-age relation to a bias due to incompleteness in source detection. 

 {In Fig.~\ref{f:cmdmods} the PMS evolutionary models constructed with the 
{\sl Frascati Raphson Newton Evolutionary Code}  \citep[{\sc franec},][]{chieffi89, 
deglinnocenti08} are shown overlaid on the optical CMD of the young LMC cluster 
LH~95 from {\sl ACS} photometry (shown also in Fig.~\ref{f:lh95-cmd} {\sl right}). Isochrones covering ages from 0.5 to 100~Myr
-- with the older isochrones being progressively bluer that the younger ones -- are overplotted to
demonstrate that the observed PMS part of the CMD may include a great range in ages (left panel).
In addition, evolutionary tracks on the CMD corresponding to 0.2~M{\solar} up to 6~M{\solar} stars
-- with tracks of larger masses being bluer and brighter that those of smaller ones -- show the variety 
in stellar masses of the PMS stars in LH~95 (right panel). }


\subsection{PMS evolutionary models: From the HRD to the CMD}\label{s:obsmodl}

Theoretical models computed for the PMS stellar evolution 
are generally expressed in terms of
physical quantities, i.e., the effective temperature and the total
bolometric luminosity, describing the evolution of stars in the HR
diagram. However, one of the primary aims of stellar evolution theory is
the explanation of the observed photometric data of stars in order to
extract their masses and ages from their magnitudes and colors, and
therefore a conversion between physical and observable quantities of the
models is required. Such a conversion is specifically described by
\cite{siess00}, who include the transformation between $T_{\rm eff}$
and $L$ to colors and magnitudes in the $UBVRI$ Cousins system and
$JHKL$ infrared bands for their models. These authors use simple
relations $T_{\rm eff}$ versus color, as well as bolometric corrections
from either \cite{siess97} or \cite{kenyon95}. These conversion tables,
derived from observations of stellar clusters, are valid for solar
metallicity dwarf stars, but the discrepancies can be significant when
one deals with populations with different stellar parameters. In
particular, the age of a star is related to the stellar radius and
therefore to the surface gravity, which introduces differences in the
spectral behavior, and therefore a population of different ages or
metallicities may require different conversion relations. This issue can
be important for e.g. cool M-type stars, for which the broad molecular
absorption bands dominate the optical spectra, affecting integrated
colors and color corrections. 

A more thorough method to analyze this issue is to make use of synthetic
atmosphere models, performing photometry directly on synthetic spectra.
\cite{girardi02} applied this method for their evolutionary models for
evolved populations, which were converted in several photometric systems
using a grid of atmosphere models described by three parameters
($[M/H]$, $\log{T_{\rm eff}}$, $\log{g}$). The choice of a reliable 
atmosphere grid is critical for cool stars, 
since their spectral energy distributions (SEDs) are dominated by 
broad molecular bands. These spectral features are gravity 
dependent, and since the stellar surface gravity varies during PMS contraction, 
optical colors of young PMS stars are age-dependent. This method has 
been applied for stars with $M<1$~M$_{\odot}$ and brown dwarfs by
\cite{baraffe98,baraffe01}. Concerning the MCs, there are only two 
published studies of such calculations, by \cite{cignoni09} for the 
SMC and by \cite{dario09} for the LMC.

In their conversion of the theoretical PMS models of \cite{siess00} into the 
observable plane \cite{dario09} 
followed an approach similar to that of \cite{girardi02}. These authors utilized the 
{\sc NextGen} \citep{nextgen} synthetic spectra to convert the theoretical 
models into colors and magnitudes, extended with the \citet{kurucz93} grid 
for the highest temperatures ($T_{\rm eff}>8000$~K). Observational models 
(both tracks and isochrones) were constructed for four assumed metallicities 
and several photometric systems, including that of {\sl ACS}  \citep[see][for a 
detailed description]{dario09}. \cite{cignoni09}  converted the {\sc franec} 
 evolutionary models  \citep{deglinnocenti08} into the observable plane, 
using the transformations for the {\sl HST} {\sc Vegamag} 
photometric system, calculated by \cite{origlia00}. In a subsequent study, 
for the conversion of the {\sc franec} grid of models \cite{dario10} applied 
the same method with  \cite{dario09}, utilizing the {\sc ames}  
family of synthetic spectra with updated opacities \citep{allardAMES}.  For these calculations, opacity 
tables are taken from \cite{ferguson05} for $\log{T_{\rm eff}} < 4.5 $ and \cite{iglesias96} 
for higher temperatures. The equation of state (EOS) is described in \cite{rogers96}. 
Both opacity tables and EOS are calculated for a heavy elements mixture equal 
to the solar mixture of \cite{asplund05}. In Fig.~\ref{f:cmdmods} a selected
sample of the evolutionary stages covered by these calculations are shown 
overlaid the CMD of LH~95.

\subsection{CMD-broadening of the positions of PMS stars}\label{s:cmdpmsbrd}

The positions of low-mass PMS stars in the CMD demonstrate a widening, which may be interpreted 
as indication of an age-spread, meaning that the star formation process may have lasted
several Myr, a time period easily detectable on the HR diagram (Fig.~\ref{f:cmdmods} - 
left). However, PMS stars 
show several peculiar characteristics due to surface activity and circumstellar accretion, 
which along with observational biases, such as unresolved binarity, crowding 
and photometric uncertainties, can cause considerable deviations of the stars 
from their theoretical positions in the observed CMD, giving false evidence of an age-spread
and wrong mass evaluation for individual stars. 
Below I describe the factors that can affect more severely the apparent CMD positions
of PMS stars.

\subsubsection{Variability}

Low-mass PMS stars exhibit variability in brightness thought to 
be caused by large starspots on the rotating stellar surface. For example, TTS  
in the Orion OB1 association are found to exhibit significant variability 
in optical bands \citep{briceno05}.  {Variable low-mass PMS stars that exhibit 
hydrogen emission lines, and frequently various forbidden emission lines,  
are classified as classical TTS (CTTS). In addition to excess IR continuum 
emission, these stars demonstrate also excess emission at UV wavelengths, 
which is believed to be consequence of disk accretion onto the stars. The accretion 
rates are typically $\sim 10^{-8}$~M{\solar}~yr$^{-1}$, relatively low compared to 
the typical infall rates during protostellar evolution \citep{lada00}.
On the other hand, variable low-mass PMS stars that typically produce 
little or no H{\alp} line emission, with H{\alp} equivalent widths less than 
10~\AA, are classified as weak-lined TTS (WTTS).}
The $V$-band variability is found to differ between CTTS and WTTS, 
showing an amplitude up to 3~mag for the former and between 0.05 and 0.6~mag
for the latter \citep{herbst94}. A variability survey of the Orion Nebula 
Cluster (ONC) in the Cousins $I$-band over 45 nights by \citet{herbst02} 
provided accurate light-curves for 767 stars between 12.5 and 16~mag, with about 
half having a peak-to-peak variation exceeding $\sim$~0.2~mag, and 10\%
exceed $\sim$~1~mag. Variability is found to correlate with infrared excess 
continuum emission, as diagnostic of circumstellar disks \citep{hillenbrand98},
in the sense that it is largely due to a combination of magnetically 
induced dark spots and a variable accretion, which becomes increasingly important 
for the higher amplitude variables. This correlation is more apparent for stars 
with $M$~\gsim~0.25~M{\solar}. On the other hand, slowly rotating stars in 
the ONC seem to have, on average, greater infrared excess emission than  their 
more rapidly rotating counterparts, providing evidence supportive to the disk-locking 
hypothesis. 

In general the canonical view of PMS variability dictates that cool spots on 
WTTS are responsible for most or all of their variations, while hot spots on CTTS
resulting from variable mass accretion from an inner disk contribute to their larger 
amplitudes and more irregular behavior. Indeed CTTS are characterized by irregular 
and extremely rapid photometric variability \citep[e.g.,][]{sherry03}. In young 
(\lsim~3~-~5~Myr) star-forming regions the ordinary fraction for CTTS is 30\% - 
50\% \citep{preibisch99}, and this fraction decreases with age so that by an age of a 
few Myr most low-mass PMS stars are WTTSs \citep{sherry04}. Under these 
circumstances, the effect of variability to the observed CMD broadening of PMS 
stars should become increasingly weaker with age, since these stars would be mostly 
WTTS. A relation between peak-to-peak variations in optical bands is derived for 
the Orion OB1b subassociation to be $\Delta I = \alpha \Delta V$ with $\alpha$ 
varying from 0.39 to 0.88 \citep{sherry03}.

A correlation  between rotation rate and position 
in the CMD is recently discovered for PMS stars in the Galactic associations Cepheus OB3b, 
NGC~2264, NGC~2362 and the ONC by \cite{littlefair11}. These authors found that 
stars which lie above an empirically determined median PMS rotate more rapidly than 
stars which lie below this sequence.  {If the  position within the CMD is interpreted as
being due to genuine age spreads within a cluster, then this would imply that the most rapidly rotating stars 
in an association are the youngest, and hence those with the highest likelihood of ongoing 
accretion.} Such a result, however, is in conflict with the existing picture of angular momentum 
evolution, according to which the stars are braked effectively by their accretion disks until the 
disk disperses. Instead, \cite{littlefair11} argue that, for a given association 
of young stars, position within the CMD is primarily a function of accretion history,
which can lead to spreads in radii and luminosity matching those observed.   

\subsubsection{Differential extinction}

Extinction, the absorption and scattering  of the starlight by the interstellar 
matter (ISM), does not appear to be uniform in star-forming regions.
As the ISM in these regions is clumpy, so is extinction, which thus is  
characterized as differential.  {Numerous studies have quantified the effect in
star-forming regions of the Milky Way. Indicatively, \cite{riaz11} find in NGC~6823 
in the Vulpecula OB1 association a significant differential reddening, and a bimodal 
distribution for $A_{\rm V}$, with a peak at $\sim$~3~mag and a broader peak at 
$\sim$~10~mag. 
\cite{pang11} also find significant differential reddening across the area centered on 
the HD~97950 star cluster in the giant {\hii} region NGC 3603, as the result of stellar 
radiation and winds interacting with an inhomogeneous dusty local ISM. 
They observe an increase from $A_{\rm V} \simeq$~4.5~mag to $\sim$~7~mag 
while moving from the central cavity to either the north or south at a distance of 2 pc 
from the cluster. These authors also note that differential reddening could be one of 
the causes why the age spread among PMS stars in HD~97950 appears to be as large 
as up to 10~Myr.
In the MCs, extinction is typically lower than in the Galaxy, and so is its variation. 
For example \cite{demarchi11} quantify the total extinction toward massive MS stars 
younger than $\sim$~3~Myr in 30~Dor to be in the range $1.3 < A_{\rm V}/{\rm mag} < 2.1$.
\cite{gouliermis11} in four LMC star-forming regions find an even smaller $A_{\rm V}$ 
variability with a maximum value between 0.8 and 1.5~mag.}
Nevertheless, differential extinction can naturally dislocate PMS stars from their original
CMD positions in a non-uniform manner, and therefore it can be an important 
factor that affects the broadening of the positions of PMS stars in the CMD. 
Comparisons of PMS stars detected in \hii\  regions of the MCs with reddening maps 
show that indeed the PMS stars located in the more obscured areas seem 
to occupy preferentially redder loci in the CMD \citep{sabbi07, hennekemper08}.

\subsubsection{Circumstellar disks: Irradiation and accretion}

Emission above stellar photospheric levels in PMS stars is attributed to dust and gas 
heating processes by a combination of irradiation and accretion associated with 
circumstellar disks. Optical and near-IR imaging surveys 
provide efficient means for diagnosing the presence of such disks, and the complete 
appreciation of our ability to detect them requires an understanding of the  
expected excesses over stellar photospheric flux levels. A combination of 
near-IR and optical databases assembled for PMS stars in the ONC 
showed that as a function of $V-I_{\rm c}$ color, the mean value of near-IR excess, 
expressed in $\Delta(I_{\rm c} - K)$, generally rises from the bluest stars to a 
peak around $V-I_{\rm c}$~=~1.75 and then declines toward redder colors. 
Earlier type, higher mass stars have smaller mean 2$\mu m$ excesses than 
the average star, while later type, lower mass stars (1~\lsim~$V-I_{\rm c}$~\lsim~2.5) 
have mean near-IR excesses distributed with mass roughly as expected from 
the disk models, i.e., stars of lower masses appear to have smaller excesses
\citep{hillenbrand98}. This dependence of excess fluxes due to circumstellar 
disks to the brightness of the PMS star suggests that the corresponding dislocation 
of PMS stars in the CMD is variable and not constant for the complete observed samples.
 {Mid- and far-IR excess emission from dusty circumstellar disks and envelopes is also 
found with the {\sl Spitzer Space Telescope} around PMS stars and YSOs in the Galaxy
\citep[e.g.,][]{povich11}, and the MCs \citep[e.g.,][]{whitney08, gruendlchu}.}

In the optical, there are two components contributing to excess due to accretion; optically 
thin emission is generated in the infalling flow, while optically thick emission comes from 
photospheric heating below the shock \citep{calvetgullbring98}. The optically thin emission 
is line-dominated similar to that of \hii\  regions. As a consequence bright Balmer emission 
is commonly used to derive mass accretion rates for PMS stars \citep[see, e.g.,][for recent results 
in the LMC and SMC respectively]{demarchi2010, demarchi2011}. Optically thick emission due to 
accretion dominates in $V$ and $I$, 
producing the phenomenon of {\sl veiling}, which strongly affects also shorter wavelengths, i.e., 
$U$ and $B$ \citep{mass04}. Veiling, as well as the existence of 
surrounding dusty reflection nebulae, can alter the $V-I$ colors of PMS stars, locating them thus 
closer or on the MS \citep{guarcello10}, challenging the suggestion that stars with H{$\alpha$} 
excess close to the MS are old PMS stars \citep[e.g.,][see also discussion in \S~\ref{s:pmsacc}]{demarchi2010}.

Flux excess is expected to vary with time due to both variations 
in the mass accretion, and stellar rotation. The
age of the population plays also important role, as more evolved young clusters have 
partially dispersed the circumstellar disks of their PMS low-mass stars and halt accretion 
in a significant fraction of them \citep{sicilia-aguilar2006}. In these clusters, 
only a small fraction of the PMS stars fluxes may be scattered by circumstellar 
material \citep[e.g.,][]{kraus-hillenbrand09}. 


\subsubsection{Unresolved binarity}

Simulations of the PMS population with $24\leq V \leq 27$~mag and 
$1.0 \leq V-I \leq 2.2$~mag in $\sigma$~Orionis cluster suggest that 
binaries shift the center of the PMS locus to a CMD position brighter 
and redder than the center of the PMS locus for single stars \citep{sherry04}.
In many young clusters in the Milky Way, a clear binary sequence
is observed to lie $\simeq$~0.75 magnitudes above the single star main-sequence
\citep{tout91}, and this is almost equal to the brightness shift that a
single star will suffer toward brighter magnitudes, if it is an
unresolved binary system with equal mass components \citep{debruijne01}.
Naturally, the overall effect of unresolved binarity to the PMS 
CMD positions depends on the binary fraction in the cluster. In principle, 
early-type stars are known to have both higher binary fraction and higher 
mass ratio than those of late spectral types \citep[e.g.,][]{zinneckerARAA07}.
Galactic OB associations show a binary fraction that does not differ 
significantly from the field \citep{mathieu94}, which is found for 
G dwarfs to be around $\sim$~58\% \citep{duquennoy91}. For M-type
stars two thrids of the investigated populations seem to have been 
born as single stars, with a binary fraction in this spectral type
of about 31\% \citep{lada06}. The well-studied ONC is also found to be consistent with the field 
binary fraction \citep[e.g.,][]{prosser94, petr98, kohler06}, 
with only 30\% of the stars in $\sigma$~Ori cluster expected to have unresolved 
binary companions \citep{sherry04}. The mass ratio of PMS 
binaries is found to follow a comparatively flat distribution for 
$M_2/M_1 \geq 0.2$ \citep{woitas01}.

\subsubsection{Source confusion and photometric accuracy}

The spatial distribution of PMS stars along the line of site can play an important role in
dislocating them in the CMD, producing a broadening of their positions in star-forming
regions of the Galaxy \citep{sherry03}. The significance of this phenomenon depends  
naturally on the crowding of sources in the observed cluster, which demonstrates itself 
as source confusion in the photometry. Crowded regions produce higher confusion in 
the detection of stars and thus broader CMD widening in the PMS positions.

Photometric accuracy depends on brightness, being higher for the 
brighter stars. Therefore, photometric uncertainties in both 
magnitudes and colors for the faint PMS stars naturally contribute to the 
broadening in their CMD positions, which then would be wider for the fainter stars. 
While, however,  photometric errors provide a reasonable explanation for the observed 
CMD widening of PMS stars, they cannot be considered a major factor, as in all cases of 
young clusters and associations observed with {\sl HST} the CMD broadening 
is proved to be much wider than the mean photometric errors per magnitude 
range \citep[see e.g.,][]{nota06, gouliermis07b,vallenari10}.

\section{Disentangling the masses and ages of PMS stars from their optical CMD}\label{s:mod2dat}

Considering the aforementioned physical and methodological factors that 
can alter the disentangling of the true nature of PMS stars from their 
positions in the optical CMD alone, one understands that the interpretation
of the observed colors and magnitudes for the extraction of the basic parameters
of the stars, i.e., their age and mass, is a  difficult task. As discussed earlier, while 
methods developed for the detection of TTS in the Galaxy allow the accurate 
determination of the characteristics of these stars, in the case of PMS stars in 
the MCs one must rely on space imaging alone. As a consequence, new treatments 
that can take advantage of the rich but single-epoch {\sl ACS} photometry of star-forming 
regions in the MCs are developed. These methods are based on the statistical/probabilistic 
extraction of the parameters of the observed PMS stars through modeling of the 
observed CMDs and the application of populations synthesis techniques.
There are only two examples of such methods for the MCs, one developed by \cite{cignoni09} 
for the SMC star-forming region NGC~346/N66, and that by \cite{dario10} for the 
LMC cluster LH~95. In both cases {\sl ACS} imaging provides rich samples of PMS stars, 
which are absolutely necessary for the use of good number statistics in these treatments. 
Nevertheless, the success of these investigations depends further on the PMS 
evolutionary models and their transformation from the theoretical H-R plane ($T_{\rm eff}$, $L$) to the observable 
C-M plane  (e.g., $V-I$, $I$) for the photometric filter system of  ACS and the metallicity of the MCs, 
as described in \S~\ref{s:obsmodl}. 



\subsection{Modeling the CMD-broadening of PMS stars}

The observed spread in brightness at a given color for PMS stars in the MCs 
can be explained in terms of various astrophysical sources of scatter 
(discussed in \S~\ref{s:cmdpmsbrd}), and therefore the broad age coverage that 
fit the observed CMDs (as shown in Fig.~\ref{f:cmdmods}, {\sl left}) does not necessarily 
correspond to any real age spread among PMS stars. It may well be 
the result of biases introduced by observational constraints, i.e., photometric 
accuracy and confusion, and the physical characteristics of these stars, such 
as variability, binarity, and circumstellar extinction. Indeed, in star-forming 
regions of  the Milky Way, whilst there probably is a true luminosity dispersion 
in the observed CMDs, there is little evidence to support age spreads larger 
than a few Myr \citep[see, e.g.,][]{jeffries11}.  {\cite{burningham05} investigated
the role of photometric variability in causing the apparent age spreads observed in
the CMDs of Galactic OB associations and they found that the combination of 
binarity, photometric uncertainty and variability on time-scales of few years is
not sufficient to explain the observed age spreads in either of the studied 
associations. In addition, \cite{jeffries07} estimated the radii of PMS stars in 
ONC, using their rotation periods and projected equatorial velocities, and he showed 
that the apparent age spreads in the HR diagram of the ONC are associated
with  a genuine spread in stellar radii.}

In the MCs, \cite{hennekemper08} demonstrated the impact of the 
CMD scat\-te\-ring factors to the interpretation of the observed 
CMDs in the SMC star-forming region NGC~346/N66 by 
constructing a simple toy-model of the influence of some of 
these factors on a ``perfect'' single-age sequence of PMS stars. These authors found that
indeed a single-age PMS population may appear in the optical CMD 
broad enough to be misinterpreted as the result of multi-epoch star 
formation. They specify the importance of reddening and binarity of 
low-mass PMS stars to this broadening, while variability
also plays an important role depending on the nature of these stars. 
Although each of these factors acting alone cannot cause the observed
widening of PMS loci in the CMD, it is their cumulative action that
leads to the broadening of the sequence of PMS stars. 
 {When applying their models to NGC 346, they find that 
an evolutionary age of about 10 Myr proved to fit
best the observed sequence of PMS stars and their broadening}. This
result is in line with previous claims that although the
location of the bright stars of NGC~346 in the CMD implies an age of
less than 5~Myr for the association, there is a spatial distinct
subgroup of evolved $\sim$~15~M{\solar} stars in the system 
with ages $\simeq$~12~Myr \citep{massey89}. Moreover, the 
simulations by \cite{hennekemper08} showed that the possibility of 
multi-epoch star formation in the region of NGC~346 cannot be excluded. 
Typical ages derived from the simulations of a double star formation event 
(assuming that 50\% of the stars are formed in each epoch) are 3 - 4 and 
10 - 12 Myr.

In subsequent studies, in order to assess the source of the brightness 
scattering of PMS stars in CMDs observed in star-forming regions of the 
MCs \cite{cignoni09, cignoni10} and \cite{dario09, dario10} make use of 
evolutionary models, as they transformed them into the C-M plane (see  
\S~\ref{s:obsmodl}) and simulate their observed CMDs with more 
sophisticated populations synthesis techniques. They derive, thus, the most 
probable stellar parameters for the detected PMS stars, according to their 
CMD positions. They utilize their results for the determination of
the age and age-spread and the low-mass IMF in star-forming 
regions of the MCs. I discuss their results, as well as those from 
other studies, in the following sections.

\section{The low-mass IMF in the Magellanic Clouds from their PMS stars}\label{s:imflh95}

The IMF of a young star cluster, i.e., the distribution of its stars according to
their mass at the time of their formation, is generally parameterized as 
\begin{equation} \xi(M)~\textrm{d}M\propto M^{-(1+x)}~~,\label{eq:imfdef}
\end{equation}  described by a series of power-laws with exponents
changing in different mass ranges \citep[e.g.,][]{scalo86,scalo98,kroupa02}. 
Reference value of the single power-law index of the IMF for stars with 
0.4~\lsim~$M$/M{\solar}~\lsim~10 in the Solar neighborhood is that 
established by \cite{salpeter}, $x=1.35$. Considering that the stellar IMF 
provides a deep insight into the star formation process, its universality,
or its dependence to environmental conditions is of critical importance
\citep{bastian10}. The investigation of whether the IMF is universal or not
requires data from other galaxies, where stars can be fairly resolved, and
the MCs are the best environments for such a study. However, while the 
high-mass IMF in the MCs is well known not to vary significantly from that 
in the Galaxy \citep{massey03}, the lack of deep data did not allow a thorough 
investigation of the low-mass IMF in the MCs, until the discovery of rich 
samples of low-mass PMS stars with {\sl HST}. 

\subsection{The low-mass stellar IMF in the Magellanic Clouds}\label{s:imfissues}

Once the stellar masses in a complete observed sample are established, the 
construction of their IMF is in principle straightforward. There are, though, three 
issues, which should be taken into account for the construction of the IMF of 
PMS populations in the MCs. (1) {\sl Field decontamination of the stellar sample}. 
The general field of the MCs is consisted mainly by evolved MS, 
sub-giants, and red giant branch stars \citep[e.g.,][]{castro01, 
smecker-hane02, javiel05, sabbi09}, and therefore it is completely 
different from star-forming regions, which comprise    
hot MS and cool PMS stars. Under these circumstances the field 
subtraction from the observed stellar samples 
can be achieved fairly well. 
(2) {\sl Fitting the IMF}. Typical linear
regression methods, used for obtaining the functional form of an
observed IMF, are based on the assumption that the measurement uncertainty
associated to each mass-bin follows a Gaussian distribution. However,  
the uncertainty in the number of counts within a mass-bin is
the overall effect of both the Poissonian error that naturally comes from the
counting process, and the uncertainty that arises from the field
subtraction. Although the first can be well
approximated with a Gaussian for large numbers, the latter depends on the
CMD positions of stars in relation to the evolutionary  tracks.  
(3) {\sl Correction for incompleteness}. In the
low-mass regime corrections for the incompleteness of the stellar sample
are required to estimate the actual number of stars. However, given the
fact that the low-mass evolutionary tracks are very close to each other 
in the CMD (Fig.~\ref{f:cmdmods}) and considering the large variation 
of completeness in both magnitudes and colors, this correction is not 
the same for all stars counted in a considered mass-bin.

\begin{figure}[t!]
\centerline{\includegraphics[clip=true,width=1.\columnwidth]{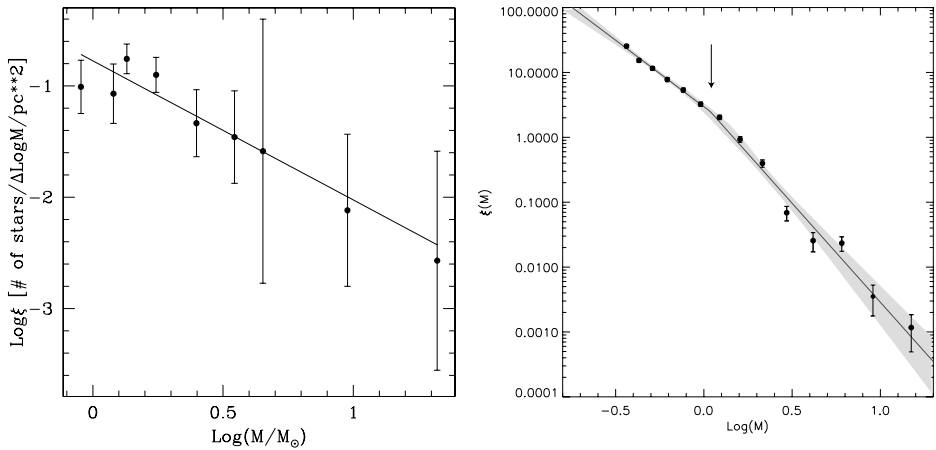}}
\caption{The IMF in the MCs. {\sl Left}: The IMF constructed for the PMS stars of the young 
SMC cluster NGC~602 \citep{cignoni09}, reproduced by permission of the AAS. The shallowness 
of the observations did not allow the construction of the IMF for the sub-solar 
PMS populations, because of insufficient completeness in their detection. This IMF
is well represented by a single power law with a slope comparable to that found by 
\cite{salpeter}, \cite{scalo98} or \cite{kroupa02} for masses larger than 1~M{\solar}.  {\sl Right}: 
The stellar IMF of the young LMC cluster LH~95 derived from its 
PMS population detected with {\sl ACS} imaging \citep{dario09}. The deeper observations of LH~95 
allowed to extend the stellar IMF to the 
sub-solar regime. The knee of this IMF at $\sim$1~M{\solar}, indicated by the arrow, 
and its multi-power law slopes are found to be comparable to the Galactic IMF 
\citep[e.g.,][]{kroupa02}. } \label{f:imf}
\end{figure}

Recently, the sensitivity and resolving efficiency of {\sl HST} imaging provided sufficient 
numbers of PMS stars in the MCs for the construction of their low-mass stellar 
IMF. Nevertheless, the required sensitivity is so high that even among the few 
available investigations with {\sl ACS} photometry, all but one are well 
limited to stars with masses \gsim~1~M{\solar} due to insufficient completeness 
in the detected stellar samples. These investigations, as well as those based
on WFPC2 imaging, agree that the PMS IMF down to this mass-limit in both the 
LMC and the SMC can be well represented by a Salpeter power-law. In particular, 
\cite{gouliermis06a} established the low-mass stellar IMF of the young LMC 
cluster LH~52 from PMS stars detected with WFPC2 imaging. They find an IMF
slope of $x \approx 1.26$ in the mass range 0.8~-~1.4~M{\solar}. The low-mass 
IMF of the bright star-forming region NGC~346/N66 in the SMC is determined by 
\cite{sabbi08} from {\sl ACS} imaging of its PMS population. These authors find a 
stellar IMF slope of $x = 1.43 \pm 0.18$ in the mass range 0.8~-~60~M{\solar}. 
This slope changes, as a function of the radial distance from the center of 
NGC~346, indicating, according to these authors, primordial mass segregation. The IMF of the young 
SMC cluster NGC~602 is addressed with photometry of {\sl ACS} images by
\cite{schmalzl08} and \cite{cignoni09}, the latter shown in Fig.~\ref{f:imf} 
({\sl left}). The results of both studies are
in agreement, with the first deriving an IMF slope of $x = 1.2 \pm 0.2$ for 
1 \lsim $M$/M{\solar} \lsim 45, and the second $x = 1.25 \pm 0.22$. 
More recently \cite{vallenari10} using imaging from five {\sl ACS}/WFC fields
provide the IMF of clusters and associations in the bright \hii\ 
complex N11 in the LMC. These authors find that the low-mass IMF slope 
varies between $x \simeq 1.0$ and $x \simeq 2.0$ for the PMS populations 
in the sub-clusters of the stellar associations LH~9, LH~10, and LH~13. 
All aforementioned investigations note that the IMF slope becomes steeper 
if the stellar samples are corrected for binarity.

%

\subsection{The sub-solar stellar IMF in the Magellanic Clouds}\label{s:subsolimf}

Concerning the IMF of PMS stars with masses \lsim~1~M{\solar} in the MCs,
\cite{liu09a} used imaging obtained with both {\sl WFPC2} and the Space 
Telescope Imaging Spectrograph ({\sl STIS}) to derive the sub-solar IMF of the
compact LMC cluster NGC~1818. They found that this IMF could be well 
approximated by both a multi-power law function \citep[e.g.,][]{kroupa01}  and a 
lognormal distribution \citep[e.g.,][]{chabrier03}. They extended their analysis
to five more clusters in the LMC with the use of the same observational material
and they found that their IMF down to the sub-solar regime is not significantly 
different from the IMF in the solar neighborhood \citep{liu09b}. In these studies the 
sub-solar PMS populations are retrieved from the photometry obtained with {\sl STIS}  
through only one filter, the F28$\times$50LP (central wavelength $\lambda_{\rm c}$ = 
7230 \AA), and therefore with no color information available for these stars. 
From comparison between evolutionary models for MS \citep{girardi02} and 
PMS \citep{baraffe98} stars these authors assess the percentage of the stars detected 
by {\sl STIS} that may be PMS stars.  The lack of any color information about these stars 
naturally introduces important uncertainties in the determination of their masses and IMF, 
which cannot be established from their HR positions. The masses of 
the {\sl STIS} sources were determined from a mass-luminosity 
relation derived from the \cite{baraffe98} PMS isochrone for the assumed age of
the cluster (Liu, private communication). Moreover, considering the age of all these  
clusters, being $\tau \gsim 40$~Myr, dynamical evolution of the clusters may have 
affected significantly the actual numbers of low-mass stars measured for the construction of their 
IMFs \citep{allison10}.

The only study, which addresses the sub-solar stellar IMF in the MCs with the use of 
both magnitudes and colors for low-mass PMS stars in a young cluster is that 
by \cite{dario09}. These authors utilize the deep photometry of the LMC star-forming 
region N64 \citep{henize56}, where the young cluster LH~95 \citep{luckehodge70} is 
embedded, imaged with {\sl ACS}/WFC \citep{gouliermis07b}. A value of mass to each 
PMS star in LH~95 is been assigned by interpolating between the PMS evolutionary 
tracks, based on the observational PMS evolutionary models for the {\sl ACS} photometric 
system and the average LMC metallicity, developed by these authors (see 
\S~\ref{s:obsmodl}). For the IMF construction, PMS stars were counted 
on the CMD in variable-sized logarithmic mass-bins, with sizes in $\log(M)$
that increase towards higher masses. Variable-sized bins yield very small 
biases, which are only weakly dependent on the number of stars, in contrast 
to uniformly binned data \citep{imfbias}. 

For the derivation of the functional form of this IMF \cite{dario09} made the 
assumption that it should be  represented by multiple power-laws, 
with the positions of the break points along the IMF, i.e., the number of
power-laws and the corresponding slopes being the free parameters of 
the fitting process. The best 
fit to the IMF was derived with the application of a Levenberg-Marquard 
non-linear least square minimization technique \citep{Levenberg44, 
Marquardt63}.  The results of the fitting algorithm for different number of 
power-laws were normalized with the so-called statistical ``F test'' based 
on the {\sl Fisher-Snedecor distribution} \citep[e.g.,][]{1965hmfw.book.....A}. 
This test showed that the IMF of LH~95 is best approximated by a two-phase 
power-law (Fig.~\ref{f:imf} {\sl right}). The slope of this IMF is found within 99\% confidence to be  
$x = 1.05^{-0.29}_{+0.20}$ for $-0.5<\log(M/{\rm M}_\odot)<0.04$ and $x = 2.05^{-0.32}_{+0.53}$ 
for $\log(M/{\rm M}_\odot)>0.04$, comparable to the average Galactic IMF \citep{kroupa02}.  
The measured 99\% confidence uncertainties, however, did not rule out variations of the IMF, 
possibly due to incomplete stellar numbers. \cite{dario09} found no significant differences in 
the shape of the overall IMF of LH~95 from that of each of its three individual PMS sub-clusters, 
suggesting that this IMF is not subject to local  variability.

\section{Timing the star formation process in the Magellanic Clouds}\label{s:timesf}

Low-mass PMS stars, as mentioned in the introduction, are excellent timers of the 
star formation process in their hosting regions, due to their long stay in this phase of 
their evolution. Therefore they are targeted by several studies for extracting the
recent star formation history (SFH) of their ambient star-forming regions, determining 
the ages of their hosting star clusters and identifying any age spreads among 
them, as indication of star formation over extended periods of time. However, as mentioned in 
\S~\ref{s:cmdposdisl}, since these studies rely on single-epoch photometry of these 
stars, they deal with problems rising by the physical characteristics and observational 
limitations that act as sources of the brightness broadening of these stars  in the 
CMDs. As a consequence, age determination of PMS stars cannot be achieved by
simple isochrone fitting to the available models, but more sophisticated populations 
synthesis techniques are developed, as described in \S~\ref{s:mod2dat}. In this section 
I discuss the results of these studies, along with others on dating the star formation
process in the MCs from their PMS populations.

\subsection{Star formation histories derived from PMS stars}\label{s:sfh}

\cite{cignoni09} presented a set of model stellar evolutionary tracks, calculated with 
the {\sc franec} evolutionary code for the typical SMC metallicity (see also  
\S~\ref{s:obsmodl}). They used these models with a stellar population synthesis 
code that takes into account a large 
range of stellar evolution phases, to derive the best estimate for the SFH in the SMC 
star-forming region NGC~602/N90. By comparing synthetic CMDs with
those constructed for the PMS stars detected with {\sl ACS}/WFC in the region, 
these authors found that the star formation 
rate has been quite high, and increased with time on a scale of tens of Myr, reaching a 
peak of (0.3~-~0.7)$\times 10^{-3}$~M{\solar}~yr$^{-1}$ in the last 2.5 Myr, comparable 
to what is found in Galactic OB associations \citep[e.g.,][]{preibisch99}.  In a subsequent
investigation \cite{cignoni11} apply the same methodology to derive the SFH of another 
star-forming region in the SMC, NGC~234/N66, based on {\sl ACS} images. They found 
that this region experienced different regimes of star formation, and in particular a 
``high-density mode'', with sub-clusters hosting both hot MS and cool PMS stars, and a 
diffuse ``low-density mode'', as indicated by the presence of low-mass PMS sub-clusters. 
According to these authors, star formation in the oldest sub-clusters started about 6 Myr 
ago with remarkable 
synchronization, it continued at a high rate (up to $2 \times 
10^{-5}$~M{\solar}~yr$^{-1}$~pc$^{-2}$) for about 3 Myr and is now progressing 
at a lower rate.

In a recent study, \cite{gouliermis11} analyzed the PMS stellar content of four star-forming 
regions located at the periphery of the super-giant shell LMC~4 (Shapley Constellation 
III) with {\sl WFPC2} imaging. While the low sensitivity of {\sl WFPC2} limited 
the detected samples of PMS stars, this investigation provided information about the star 
formation process around the shell without any age determination for the stars. Specifically, 
\cite{gouliermis11} compared the distributions of low-mass PMS stars along cross-sections
of the observed CMDs in the different regions. These authors found that these distributions 
in the regions LH~60/63/N51 and LH~72/N55 exhibit an extraordinary similarity, suggesting 
that these PMS stars share common characteristics, as well as common recent SFH. Considering 
that the regions are located at different areas of the edge of LMC 4, at distance of about 500~pc, 
this finding suggests that star formation along the whole shell may have occurred at 
the same time.

\subsection{Age determination of a cluster from its PMS stars}\label{s:agelh95}

 {In the LMC \hii\  complex N11, \cite{vallenari10} determined the ages of individual clusters
and associations from {\sl ACS}/WFC imaging using the luminosity function of the PMS 
populations.} While this approach does not take into account the uncertainties introduced
by the CMD-broadening of these stars, its results provide a uniform set of ages 
of clusters within a large LMC star-forming region, and thus allow a coherent understanding
of its recent star formation process. These authors found from the PMS population of clusters in the association 
LH~9 a prolonged formation of stars from 2 to 10~Myr, in agreement with previous spectroscopic
studies of the hot blue stellar content of the system \citep{walborn92, mokiem07}, with the 
majority of the stars being older than 6~Myr. On the other hand, they 
found that  the dominant PMS populations in the neighboring associations LH~10 and LH~13 
have ages of 2 to 3~Myr and 2 to 5~Myr respectively, again in agreement with results from 
spectroscopy of their bright stars \citep{HM00, mokiem07}. Candidate YSOs 
 with ages from 0.1 to 1~Myr are found by the same authors at the location of 
the PMS stars in the region with observations with the {\sl Spitzer Space Telescope}.
This indicates that several generations of stars are present in the region of N11.
 \cite{vallenari10} propose a star formation scenario for the region, according to which
the formation of the first generation of stars in LH~10 was triggered by supernova 
shocks in the association LH~9.

A more sophisticated method for the determination of the age of individual sub-clusters in a star-forming 
region from its PMS stars is developed by \cite{cignoni10}. These authors use the Turn-On (TOn), i.e., the 
point in the observed CMD where the PMS joins the MS, and apply a method akin to that by \cite{piskunov04}. 
According to this method it is possible to reliably identify the TOn, which in the MS luminosity function (LF) 
of the cluster appears as a peak followed by a dip, from the monitoring of the 
spatial distribution of MS stars. \cite{cignoni10}  simulated synthetic simple stellar populations 
covering a set of five different ages by using PMS evolutionary models, and through a polynomial fit to the corresponding 
model LFs they established a functional relation between age and the magnitude that corresponds to the TOn.
According to the authors this CMD analysis can provide a reliable age of extragalactic clusters, but 
it should be used with caution because of uncertainties in the comparison between the
observed and theoretical LFs, emanating from incompleteness and photometric errors, Poissonian fluctuations
in the stellar numbers, reddening, the assumed IMF, binarity and the star formation duration.
The application of this technique, which is complementary to the turnoff dating and avoids the systematic biases affecting 
the PMS phase, to three sub-clusters of the  region NGC~346/N66 showed that all sub-clusters have 
different ages, varying between 3 and 18 Myr, supporting the idea of complexity in the recent SFH 
of the region.

\cite{dario10} assessed the age of the young LMC cluster LH~95 and determined the possible 
appearance of an age spread from the CMD-broadening of its PMS stars with the development 
of a self-consistent maximum-likelihood method, especially designed for {\sl ACS} photometric data. 
With their method, which is similar to that presented by \cite{naylor-jeffries06}, these authors aimed at the 
determination of the age of the cluster accounting simultaneously for the most significant biases that 
affect the CMD positions of the PMS stars, as they are discussed in \S\ref{s:cmdposdisl}. \cite{dario10}
performed calculations of PMS evolutionary models for the LMC metallicity using the {\sc franec}
code and transformed them, as well as the \cite{siess00} family of models, into the {\sl ACS} photometric 
system (see also \S~\ref{s:obsmodl}). They converted these models to 2D probability distributions in the 
magnitude-magnitude plane, and consequently in the CMD, by applying the most 
important sources of displacement of the CMD positions of PMS stars
among those discussed in  \S\ref{s:cmdposdisl}, except of photometric errors. The assumed 
IMF for LH~95 is that derived by \cite[][see also \S~\ref{s:subsolimf}]{dario09}.

A maximum-likelihood method is then applied to derive the probability for each observed 
star to have a certain age, taking its photometric uncertainty into account, and considering  
the Gaussian distributions of the photometric errors as priors for the determination of the 
 likelihood of each star as function of age. This process of modeling 
age-dependent probability distributions and maximizing their global 
likelihood functions is applied with the use of two different sets of evolutionary models, available 
for the metallicity of the LMC, that by \citet{siess00}, and that computed by \cite{dario10}  with the 
{\sc franec} stellar evolution code \citep{chieffi89, deglinnocenti08} and for different assumed binary 
fractions. This treatment showed that the age 
determination is sensitive to the assumed evolutionary model and binary fraction; the age
of LH~95 is found to vary from about 2.8 to 4.4~Myr, depending on these factors. 
The best-fit age of the cluster LH~95 derived for a binary fraction of $f=0.5$, which was 
considered a reasonable value in the mass range of interest \citep{lada06}, 
is determined by \cite{dario10} to be between 3.9 and 3.8~Myr.

\subsection{Assessment of an age spread in PMS clusters}

Distinguishing if the observed CMD-broadening of PMS stars represents also a true age spread, 
or is the result of the intrinsic properties of these stars and their measurement uncertainties is one 
of the major tasks in our understanding of star formation in clusters of the MCs. With their method, 
\cite{dario10} investigated whether the spread of the PMS stars in the observed CMD is wider or 
not than that of the simulated 2D density distributions as they are produced by accounting for the 
sources of CMD-broadening alone. If indeed the observed CMD-broadening of PMS stars 
was wider than the synthetic derived from a single isochrone, this would indicate the presence of 
a real age spread in LH~95.  \cite{dario10}, by comparing the observed and modeled CMD 
distributions and evaluating the goodness-of-fit with a method similar to a standard $\chi^2$ 
minimization,  showed that the observed sequence of PMS stars in the CMD of LH~95 is 
broader than what predicted by the 2D models for single ages, indicating that the cluster 
probably hosts a real age spread among its PMS stars. These authors quantified this 
spread to be of the order of $\sigma_{\rm age}=1.8$~Myr (FWHM$\simeq 4.2$~Myr), 
as derived with the use of the {\sc franec} models, while it decreases to 
$\sigma_{\rm age}=1.2$~Myr (FWHM$\simeq 2.8$~Myr) if \cite{siess00} models
are considered. In both cases the uncertainty in $\sigma_{\rm age}$ is about 0.2~Myr.

\subsection{Star formation histories from accretion studies of PMS stars}\label{s:pmsacc}

\cite{demarchi2010}, based on the original study by \cite{panagia00}, developed a new 
self-consistent method to identify PMS 
objects actively undergoing mass accretion in a resolved stellar population, regardless of their age. 
The method combines {\sl HST}  broadband $V$- and $I$-equivalent photometry with 
narrowband H\alp\ imaging to identify stars with excess emission and obtain their mass 
accretion rate $\dot{M}_{\rm acc}$. The application of this method in a field around  
SN~1987A in the LMC derived a median mass accretion rate for the detected PMS stars 
of 26$\times$10$^{-8}$~M{\solar}~yr$^{-1}$, in agreement with previous  
determinations based on the U-band excess of PMS stars in the same field \citep{romaniello04}.
This study revealed a sample of PMS stars located in the CMD at the MS. The ages of all identified 
objects were determined by \cite{demarchi2010} by interpolating between the isochrones.
These authors found that the detected PMS stars cover an age-range of 1 to $\sim$50~Myr, with
the CMD locations of the older PMS stars coinciding with those of MS stars. 

More recently, in a subsequent study, \cite{demarchi11} applied the same method and identified 
about 1150 PMS stars with a strong H{\alp} excess in 30 Dor.  Comparison of their location in 
the H-R diagram with theoretical PMS evolutionary tracks revealed that about one-third of these 
objects are younger than $\sim$4~Myr, whereas the rest have ages up to $\sim$30 Myr.
These authors argue that this indicates that star formation has proceeded over an extended 
period of time, although they cannot discriminate between an extended episode and a series 
of short and frequent bursts that are not resolved in time. A bimodal age distribution of PMS stars
with H{\alp} excess, with two roughly equally numerous populations peaked respectively at  
$\sim$1~Myr and $\sim$20~Myr, is also been detected by the same team in the SMC star-forming 
region NGC~346/N66 \citep{demarchi2011} with the application of their accretion determination  
method. 

These studies, which reveal accreting `blue' PMS stars, located close or on the MS, 
provide reasonable  indications of  large age spreads among PMS stars in star-forming 
regions of both the MCs. However, the age determination based on direct isochrone fitting 
on the H-R diagram, without considering the CMD-broadening sources of PMS stars  
discussed in \S~\ref{s:cmdposdisl}, decrease considerably the accuracy of this determination. 
In particular, PMS isochrones representing ages older than $\sim$10~Myr are very close 
to each other and to the lower MS (Prada Moroni, private communication; see also 
Fig.~\ref{f:cmdmods}), and thus the distinction of a 
20~Myr old PMS population from that with age of 30, 50 (as presented by 
De~Marchi et al.), or even 100~Myr is not straightforward. 
Moreover, `blue' PMS stars in the Galaxy are suggested to be normal PMS stars with 
altered colors \citep{guarcello10}. This behavior has been previously 
documented and a possible explanation of the nature of these stars is that 
they are strong accretors, characterized by an intense veiling that 
alters their optical colors \citep[e.g.,][]{hartmann90} or their photometry might 
be contaminated by the nebula emission \citep[e.g.,][]{hillenbrand98}, or they are 
surrounded by dusty reflection nebulae \citep[e.g.,][]{damiani06}. Another possible 
explanation is that a significant fraction of  stellar radiation is scattered 
into the line of sight by the circumstellar disk \citep{throop01, robitaille06}. \cite{guarcello10} 
found by simulating synthetic CMDs that this explanation holds for disks 
observed at high inclination around PMS stars in the Eagle nebula. However, 
more recent simulations suggest that the percentage of PMS stars with  
highly inclined disks is not enough to explain the large numbers of the detected
`blue' PMS stars in NGC~346 \citep{demarchi2011}.

\section{Conclusive Remarks}\label{s:conclrem}

I present  results on the relatively unexplored field of PMS evolution in the MCs, and 
in particular the investigation of PMS stellar populations with optical space imaging. 
I specifically discuss the cases of star-forming regions 
in the MCs, which have been observed with {\sl ACS} onboard {\sl HST}.
Low-mass PMS stars remain in 
the same evolutionary stage for several tens Myr, and therefore host 
unprecedented information about the age and star formation process of 
their host cluster, as well as  its stellar IMF. Since these stars demonstrate 
peculiarities in their physical  properties due to coronal and circumstellar 
activity, their investigation requires demanding observations and their 
careful interpretation. Observational techniques, such as  
optical photometric monitoring, spectroscopy, and X-ray imaging
are essential for the decontamination of the PMS stellar 
samples in star-forming regions of the Milky Way from the fore- and 
background stars of the Galactic plane. Considering that only {\sl HST} can provide the 
appropriate combination of sensitivity  and field-of-view coverage at the MCs, 
it is practically impossible to perform repeated imaging, spectroscopy of 
large samples, or X-ray monitoring of star-forming regions in these galaxies, and 
investigators rely on single-epoch photometry alone.
There are, however,  two major advantages in studying PMS stars in the MCs, 
which allow the derivation of important results
concerning the properties of young clusters that host such stars. First, 
the observed PMS populations in the MCs can be quite 
easily distinguished from the evolved stellar populations of the general field of the 
galaxies directly on the CMD. Second, deep imaging with {\sl HST} provides 
outstanding numbers of PMS stars in star-forming regions of these 
galaxies, which are large enough for detailed statistical analyses 
of their properties, provided that the behavior of PMS stars in the optical CMD is 
sufficiently modeled.

Such statistical treatments for the derivation of the stellar IMF and age of 
young half-embedded clusters in the MCs, are discussed, 
and  important physical  sources of the CMD-spread of PMS 
stars, which affect the outcome of these treatments, are presented. In particular, 
I discuss rotational variability, accretion and differential extinction, among others, 
as very important factors that naturally dislocate the PMS stars from their original 
(theoretical) CMD positions, and which should be taken into account in the interpretation 
of the observed CMDs. Since these studies are based on imaging, 
there are three additional factors that further bias the observed CMD loci of PMS 
stars: unresolved binarity, source confusion and photometric uncertainties. 
I also stress the importance of the use of accurate PMS evolutionary models 
that are appropriately designed for the metallicity of the MCs and correctly 
translated in the photometric system of the cameras. The use
of such evolutionary tracks in combination with a detailed treatment for the 
contamination of the stellar sample by the field population can provide 
the best bias-free solution for the most probable age and IMF of young 
clusters in the MCs through their PMS populations, observed with {\sl HST}.

\begin{acknowledgements}
I kindly acknowledge financial support from the German Research Foundation (DFG) 
through grants GO~1659/1-1 and GO~1659/1-2, and the German Aerospace Center 
(DLR) through grants 50~OR~0401 and 50~OR~0908. I am benefitted by the 
collaboration with several colleagues, whom I thank for their insight and support. 
In particular, I am grateful to Th. Henning, W. Brandner and M. Gennaro (MPIA), 
N. Da~Rio and G. De~Marchi (ESA), M. 
Robberto and N. Panagia (STScI), A. E. Dolphin (Raytheon Company), and P. G. 
Prada Moroni and E. Tognelli (Universit\`{a}  di Pisa). I am also 
indebted to A. Nota, E. Sabbi and L. J. Smith (STScI) for our fruitful discussions about
star-forming regions in the MCs. Last but not least, I would like to express my
sincere thanks to the unknown referee for her/his constructive comments, which 
helped significantly to the accuracy and completeness of this review. Based on observations
made with the NASA/ESA {\em Hubble Space Telescope}, obtained 
at the Space Telescope Science Institute. STScI is operated
by the Association of Universities for Research in Astronomy, Inc. under
NASA contract NAS 5-26555. 
\end{acknowledgements}



\end{document}